\documentclass[12pt]{article}
\usepackage[english,german,french,polish]{babel}
\usepackage[T1]{fontenc}
\usepackage{amsfonts}

\begin{document}

\selectlanguage{english}

\baselineskip 0.76cm
\topmargin -0.6in
\oddsidemargin -0.1in

\let\ni=\noindent

\renewcommand{\thefootnote}{\fnsymbol{footnote}}

\pagestyle {plain}

\setcounter{page}{1}

\pagestyle{empty}

~~~

\begin{flushright}
IFT-- 04/3\end{flushright}

\vspace{0.4cm}

{\large\centerline{\bf Guesswork for Dirac and Majorana neutrino }}

{\large\centerline{\bf mass matrices{\footnote 
{Work supported in part by the Polish State Committee for Scientific Research 
(KBN), grant 2 P03B 129 24 (2003--2004).}}}}

\vspace{0.4cm}

{\centerline {\sc Wojciech Kr\'{o}likowski}}

\vspace{0.3cm}

{\centerline {\it Institute of Theoretical Physics, Warsaw University}}

{\centerline {\it Ho\.{z}a 69,~~PL--00--681 Warszawa, ~Poland}}

\vspace{0.3cm}

{\centerline{\bf Abstract}}

\vspace{0.2cm}

In the framework of seesaw mechanism with three neutrino flavors, we propose 
tentatively an efficient parametrization for the spectra of Dirac and 
righthanded Majorana neutrino mass matrices in terms of three free parameters. 
Two of them are related to (and determined by) the corresponding parameters 
introduced previously for the mass spectra of charged leptons and up and down 
quarks. The third is determined from the experimental estimate of solar 
$\Delta m^2_{21}$. Then, the atmospheric $\Delta m^2_{32}$ is {\it predicted} 
close to its experimental estimation. With the use of these three parameters 
all light active-neutrino masses $ m_1 < m_2 < m_3$ and heavy sterile-neutrino 
masses $ M_1 < M_2 < M_3$ are readily evaluated. The latter turn out much more 
{\it hierarchical} than the former. The lightest heavy mass $ M_1$ comes out 
to be of the order $O(10^6\;{\rm GeV})$ so, it is too light to imply that the 
mechanism of baryogenesis through thermal leptogenesis might work.

\vspace{0.6cm}

\ni PACS numbers: 12.15.Ff , 12.15.Hh , 14.60.Pq .

\vspace{0.6cm}

\ni January 2004  

\vfill\eject

~~~
\pagestyle {plain}

\setcounter{page}{1}

\vspace{0.2cm}

Some time ago we proposed for charged leptons $e_i=e^- , \mu^- , \tau^- $
an efficient empirical mass formula [1]

\begin{equation}
m_{e_i} = \rho_i \mu^{(e)} 
\left(N^2_i + \frac{\varepsilon^{(e)} - 1}{N^2_i} \right) \,,
\end{equation}

\ni where 

\begin{equation}
N_i = 1,3,5 \;,
\end{equation}

\ni and

\begin{equation}
\rho_i = \frac{1}{29} \,,\,\frac{4}{29} \,,\,\frac{24}{29} 
\end{equation}

\ni ($\sum_i \rho_i = 1$), while $\mu^{(e)} > 0$ and $\varepsilon^{(e)} > 0$ 
are constants. In fact, with the experimental values $m_e = 0.510999$ MeV and 
$m_\mu = 105.658$ MeV as an input, the formula (1) rewritten explicitly as

\begin{equation}
m_e = \frac{\mu^{(e)}}{29} \varepsilon^{(e)} \;,\;  
m_\mu = \frac{\mu^{(e)}}{29} \frac{4}{9} (80 +  \varepsilon^{(e)}) \;,\;  
m_\tau = \frac{\mu^{(e)}}{29} \frac{24}{25} (624 + \varepsilon^{(e)})
\end{equation}

\ni leads to the {\it prediction}

\begin{equation}
m_\tau = \frac{6}{125}(351 m_\mu - 136 m_e) = 1776.80 \;{\rm MeV}
\end{equation}

\ni and also determines both constants

\begin{equation}
\mu^{(e)} = \frac{29(9m_\mu - 4m_e)}{320} = 85.9924 \;{\rm MeV} \;,\; 
\varepsilon^{(e)} = \frac{320 m_e}{9m_\mu - 4m_e} = 0.172329 \,.
\end{equation}

\ni The prediction (5) lies really close to the experimental value 
$m^{\rm exp}_\tau = 1776.99^{+0.29}_{-0.26}$ MeV [2]. Though the formula (1) 
has essentially the empirical character, there is a speculative background for 
it based on a K\"{a}hler-like extension of Dirac equation which the interested 
reader may find in Ref. [1]. In particular, the numbers $N_i$ and 
$\rho_i \; (i=1,2,3)$ given in Eqs. (2) and (3) are interpreted there.

The charged-lepton mass formula [1] was recently extended to up and down 
quarks, $ u_i = u , c , t $ and $ d_i = d , s , b $, by introducing an 
additional term for the third quark generation, leading to [3]

\begin{equation}
m_{u_i} = \rho_i \mu^{(u)} \left(N^2_i + 
\frac{\varepsilon^{(u)} - 1}{N^2_i} + \delta_{i\,3}\beta^{(u)} \right) 
\end{equation}

\ni and

\begin{equation}
m_{d_i} = \rho_i \mu^{(d)} \left(N^2_i + 
\frac{\varepsilon^{(d)} - 1}{N^2_i} + \delta_{i\,3}\beta^{(d)} \right) \,,
\end{equation}

\ni where $N_i$ and $\rho_i $ are given as before in Eqs. (2) and (3), 
while $\mu^{(u,d)} > 0$,  $\varepsilon^{(u,d)} > 0$ and $\beta^{(u,d)} > 0$ 
are constants. It is seen that {\it a priori} Eqs. (7) and (8) cannot give us 
any mass predictions, since there are six quark masses and six free parameters.
However, the latter are uniquely determined. In fact, assuming for quark 
masses their mean experimental estimates [2]

\begin{equation}
m_u \sim 3\;{\rm MeV}\;\;,\;\;m_c 
\sim 1.2\;{\rm GeV}\;\;,\;\;m_t \sim 174\;{\rm GeV}
\end{equation}

\ni and

\begin{equation}
m_d \sim 6.75\;{\rm MeV}\;\;,\;\;m_s \sim 118\;{\rm MeV}\;\;,\;\;
m_b \sim 4.25\;{\rm GeV}\;,
\end{equation}

\ni one can calculate

\begin{equation}
m_{t,b} = \frac{6}{125}(351 m_{c,s} - 136 m_{u,d}) + \frac{24}{29} 
\mu^{(u,d)} \beta^{(u,d)} \sim \left\{\begin{array}{r} 20\;\:\:\; + 
\:0.81\:\, \beta^{(u)} \\ 1.94 + 0.078\, \beta^{(d)} \end{array}\right\} 
\;{\rm GeV}
\end{equation}

\ni and

\begin{equation}
\mu^{(u,d)} = \frac{29(9m_{c,s} - 4m_{u,d})}{320} \sim 
\left\{\begin{array}{r} 978\;\; \\ 93.8  \end{array}\right\}  \,{\rm MeV} 
\,,\, \varepsilon^{(u,d)} = \frac{320 m_{u,d}}{9m_{c,s} - 4m_{u,d}} \sim 
\left\{\begin{array}{l} 0.0890 \\ 2.09  \end{array}\right\} \,.
\end{equation}

\ni From Eqs. (11) it follows that

\begin{equation}
\beta^{(u)} \sim 190 \;,\; \beta^{(d)} \sim 30 \;, 
\end{equation}

\ni thus 

\begin{equation}
\frac{\beta^{(u)}}{\beta^{(d)}} \sim 6.3 \;. 
\end{equation}

\ni It may be interesting to note that the experimental ratio (14) is nicely 
reproduced by the ansatz

\begin{equation}
\beta^{(u,d)} \propto (3B + Q^{(u,d)})^2 = \left\{\begin{array}{r} 25/9 \\ 
4/9  \end{array}\right\} \,,
\end{equation}

\ni where $B\! = \!1/3$ and 
$Q^{(u,d)}\! = \!\begin{footnotesize}\! \left\{\begin{array}{r} 
\!2/3 \\ \!-1/3 \end{array} \right\} \end{footnotesize}$ are the baryon 
number and electric charge of quarks. Then, 
${\beta^{(u)}}/{\beta^{(d)}}\! = 6.25$. Note also that the analogical 
constant for charged leptons vanishes, $\beta^{(e)} \propto  
(L + Q^{(e)})^2 = 0$, where $L=1$ and $Q^{(e)}=-1$ are their lepton number 
and electric charge ($F = 3B+L$ is the fermion number as defined for quarks 
and charged leptons).

In the present note we discuss the mass spectrum of three active (lefthanded) 
neutrinos $\nu_{\alpha L} = \nu_{e L}, \nu_{\mu L}, \nu_{\tau L}$ related to 
their mass states $\nu_{i L} = \nu_{1 L}, \nu_{2 L}, \nu_{3 L}$ through the 
unitary mixing transformation 

\begin{equation}
\nu_{\alpha L}  = \sum_i U_{\alpha i}\, \nu_{i L}\;,
\end{equation}

\ni where the neutrino mixing matrix $U = \left( U_{\alpha i}\right)$ is 
experimentally consistent with the bilarge form [4]

\begin{equation}
U = \left( \begin{array}{ccc} c_{12} & s_{12} & 0 \\ - \frac{1}{\sqrt2} 
s_{12} & \frac{1}{\sqrt2} c_{12} & \frac{1}{\sqrt2}  \\ 
\frac{1}{\sqrt2} s_{12} & -\frac{1}{\sqrt2} c_{12} & \frac{1}{\sqrt2}  
\end{array} \right)\,, 
\end{equation}

\ni where $c_{12} = \cos \theta_{12}$ and $s_{12} = \sin \theta_{12}$ with 
large $\theta_{12} \sim 33^\circ $, while $c_{23} = 
\cos \theta_{23} = 1/\sqrt2$ and $s_{23} = \sin \theta_{23} = 1/\sqrt2$ with 
maximal $\theta_{23} = 45^\circ $. In Eq. (17) the matrix element 
$U_{e3} = s_{13}\exp(-i\delta)$ is neglected, where 
$s_{13} = \sin \theta_{13}$ with $s^2_{13} < 0.03$. Three sterile 
(righthanded) neutrinos $\nu_{\alpha R} = \nu_{e R}, \nu_{\mu R}, 
\nu_{\tau R}$ and their mass states $\nu_{i R} = \nu_{1 R}, \nu_{2 R}, 
\nu_{3 R}$ appear as a background.  

Our starting point will be the generic $6\times 6$ mass matrix

\begin{equation}
\left( \begin{array}{cc} 0 & M^{(D)} \\ M^{(D)\,T} & M^{(R)} \end{array} 
\right) 
\end{equation}

\ni (in the basis of active $\nu_{\alpha L}$ and sterile $\nu_{\alpha R}$), 
involving Dirac and righthanded Majorana $3\times 3$ mass matrices, $M^{(D)}$ 
and $M^{(R)}$. Accepting the familiar seesaw mechanism [5] we will use for 
active neutrinos the effective Majorana $3\times 3$ mass matrix of the form

\begin{equation}
M^{(\nu)} = M^{(D)} {M^{(R)}}^{-1} M^{(D)\,T} \,,
\end{equation}

\ni where $M^{(R)}$ is assumed to dominate over $M^{(D)}$. For the 
eigenvalues $m^{(D)}_{\nu_i} = m^{(D)}_{\nu_1},$ $ m^{(D)}_{\nu_2}, 
m^{(D)}_{\nu_3}$ of the Dirac neutrino mass matrix $M^{(D)}$ we will accept 
tentatively the formula of the same type as Eq. (1) for charged leptons,

\begin{equation}
m^{(D)}_{\nu_i} =  \rho_i \mu^{(\nu)} \left(N^2_i + \frac{\varepsilon^{(\nu)} 
- 1}{N^2_i} \right) \,,
\end{equation}

\ni where $\mu^{(\nu)} >0$ and $\varepsilon^{(\nu)} >0$ are new constants.

In the flavor representation, where the charged-lepton mass matrix $M^{(e)}$ 
is diagonal, the neutrino mixing matrix $U$ is at the same time the neutrino 
diagonalizing matrix,

\begin{equation}
U^\dagger M^{(\nu)} U = {\rm diag} (m_{\nu_1} , m_{\nu_2} , m_{\nu_3}) \,.
\end{equation}

\ni Here, for simplicity, $M^{(\nu)*} = M^{(\nu)}$ and $U^* = U$ 
[as in Eq. (17)]. In the case of seesaw form (19) of $M^{(\nu)}$, it seems 
natural to conjecture that $U$ is also the diagonalizing matrix for the Dirac 
neutrino mass matrix $M^{(D)}$,

\begin{equation}
U^\dagger M^{(D)} U = {\rm diag} (m^{(D)}_{\nu_1} , m^{(D)}_{\nu_2} , 
m^{(D)}_{\nu_3}) \,.
\end{equation}

\ni Notice that then the inverse ${M^{(R)}}^{-1}$ in Eq. (19) is diagonalized 
by $U$ as well, since from Eq. (19) ${M^{(R)}}^{-1} = {M^{(D)}}^{-1} 
M^{(\nu)} {M^{(D)T}}^{-1}$ (if the inverse ${M^{(D)}}^{-1}$ exists {\it i.e.}, 
all $m^{(D)}_{\nu_i} \neq 0 $) and both $M^{(\nu)}$ and ${M^{(D)}}^{-1}$ on 
the rhs are diagonalized by $U$. Hence

\begin{equation}
U^\dagger M^{(R)} U = {\rm diag} (M_{\nu_1} , M_{\nu_2} , M_{\nu_3}) \,,
\end{equation}

\ni when Eq. (22) is conjectured.

Thus, under the conjecture (22) we obtain from the seesaw form (19) of 
$M^{(\nu)}$ the following Majorana mass spectrum for light active 
(lefthanded) neutrinos $\nu_{i L}$:

\begin{equation}
m_{\nu_i} = \frac{m^{(D)2}_{\nu_i}}{M_{\nu_i}}\,
\end{equation}

\ni with $M_{\nu_i} \gg m^{(D)}_{\nu_i} \gg m_{\nu_i}$, where $M_{\nu_i}$ 
are the Majorana masses of heavy sterile (righthanded) neutrinos $\nu_{i R}$. 
Here, for simplicity, $ M^{(R)*} =  M^{(R)}$.

In order to proceed further with calculations of $ m^{(D)}_{\nu_i}$ [from 
Eq. (20)] and $ m_{\nu_i}$ [from Eq. (24)] we are forced to make some 
tentative conjectures about $ \mu^{(\nu)}$ and $ \varepsilon^{(\nu)}$ as 
well as $M_{\nu_i}$. We will propose tentatively that

\begin{equation}
\mu^{(\nu)}\;{\bf :}\; \mu^{(e)} = \mu^{(u)}\;{\bf :}\; \mu^{(d)} \;,\; 
\varepsilon^{(\nu)}\;{\bf :}\; \varepsilon^{(e)} = \varepsilon^{(u)}\;
{\bf :}\; \varepsilon^{(d)} 
\end{equation}

\ni and also

\begin{equation}
M_{\nu_i} \propto N^2_i m^{(D)} _{\nu_i}\,,
\end{equation}

\ni where $N_i = 1,3,5$ as before  in Eq. (2) [in Ref. [6] we conjectured 
tentatively that $M_{\nu_i} \propto N^2_i m_{e_i}$ instead of Eqs (26)]. 
With the use of values (6) and (12), Eqs. (25) imply that

\begin{equation}
\mu^{(\nu)} \sim 896 \;{\rm MeV}\;,\; \varepsilon^{(\nu)} 
\sim 7.35\times 10^{-3} \,.
\end{equation}

\ni Then, the mass formula (20) gives the following {\it hierarchical} 
estimates of Dirac neutrino masses:

\begin{eqnarray}
m_{\nu_1}^{(D)} & = & \frac{\mu^{(\nu)}}{29} \varepsilon^{(\nu)} 
\sim 0.227\;{\rm MeV} \ll \frac{\mu^{(\nu)}}{\mu^{(e)}} m_e \,, \nonumber \\ 
m_{\nu_2}^{(D)} & = & \frac{\mu^{(\nu)}}{29} \frac{4}{9}
\left(80 + \varepsilon^{(\nu)}\right) \sim 1.10\;{\rm GeV} 
\sim \frac{\mu^{(\nu)}}{\mu^{(e)}} m_\mu \,, \nonumber \\ 
m_{\nu_3}^{(D)} & = & \frac{\mu^{(\nu)}}{29} \frac{24}{25}
\left(624 + \varepsilon^{(\nu)}\right) \sim 18.5\;{\rm GeV} 
\sim \frac{\mu^{(\nu)}}{\mu^{(e)}} m_\tau \,. 
\end{eqnarray}

\ni The (weighted) proportionality relation (26), when applied to the 
seesaw spectrum (24), leads to

\begin{equation}
m_{\nu_i} \propto \frac{1}{N^2_i} m^{(D)}_{\nu_i}\,.
\end{equation}

\ni This shows that $m_{\nu_i}$ are less {\it hierarchical} than 
$m^{(D)}_{\nu_i}$. From Eqs. (29) we can see that 

\begin{equation}
\frac{m_{\nu_1}}{ m_{\nu_2}} = 
9 \frac{m^{(D)}_{\nu_1}}{m^{(D)}_{\nu_2}} \sim 1.86\times 10^{-3} \;,\; 
\frac{m_{\nu_2}}{ m_{\nu_3}} = \frac{25}{9} 
\frac{m^{(D)}_{\nu_2}}{m^{(D)}_{\nu_3}} \sim 0.165\,.
\end{equation}

\ni Thus, taking the experimental estimate $m^{\rm exp}_{\nu_2}\! 
=\! \sqrt{(\Delta m^2_{21})^{\rm exp}} 
\!\sim \!\sqrt{7\!\times\! 10^{-5}}\,$eV $= 8.4\times 10^{-3}$eV as an 
input, we {\it predict} from Eqs. (30) that 

\begin{equation}
m_{\nu_1} \sim 1.6\times 10^{-5}\;{\rm eV}\;,\;m_{\nu_3} \sim 
\sqrt{2.6\times 10^{-3}}\;{\rm eV}= 5.1\times 10^{-2}\;{\rm eV}\;.
\end{equation}

\ni The prediction $m_{\nu_3} \sim \sqrt{2.6\times 10^{-3}}$ eV gives 
$\Delta m^2_{32} = m^2_{\nu_3} - (m^{\rm exp}_{\nu_2})^2 
\sim 2.5\times 10^{-3} \;{\rm eV}^2$ close to the experimental estimate 
$(\Delta m^2_{32})^{\rm exp} \sim 2.5\times 10^{-3}\;{\rm eV}^2$ (the 
lower estimate $(\Delta m^2_{32})^{\rm exp} \sim 2\times 10^{-3}\;{\rm eV}^2$ 
would correspond to the lower prediction $m_{\nu_3} 
\sim \sqrt{2.1\times 10^{-3}}$ eV).

Denoting the proportionality coefficient in Eq. (26) by $\zeta$, we have

\begin{equation}
M_{\nu_i} = \zeta N^2_i m^{(D)}_{\nu_i}
\end{equation}

\ni and $m_{\nu_i} = (1/\zeta) m^{(D)}_{\nu_i}/N^2_i = (1/\zeta^2) 
M_{\nu_i}/N^4_i $ due to Eq. (29). Thus, $\zeta$ may be calculated {\it e.g.} 
from the relation

\begin{equation}
\zeta = m^{(D)}_{\nu_2}/9m_{\nu_2} \sim 1.46\times 10^{10}\,,
\end{equation}

\ni where the experimental estimate $m^{\rm exp}_{\nu_2} 
\sim \sqrt{7\times 10^{-5}}$ eV is applied. Then, using the values (28), 
we obtain from Eqs. (32) the following {\it hierarchical} estimates of 
Majorana sterile-neutrino masses:

\begin{eqnarray}
M_{\nu_1} & = & \;\;\;\zeta m^{(D)}_{\nu_1} \sim 3.3\times 10^6\,\;{\rm GeV} 
\,, \nonumber \\ 
M_{\nu_2} & = & \;\,9\zeta m^{(D)}_{\nu_2} \sim 1.4\times 10^{11}\;{\rm GeV} 
\,, \nonumber \\ 
M_{\nu_3} & = & 25\zeta m^{(D)}_{\nu_3} \sim 6.8\times 10^{12}\;{\rm GeV} \,. 
\end{eqnarray}

\ni It is seen that $m^{(D)}_{\nu_i}$ are less {\it hierarchical} than 
$M_{\nu_i}$ (and $m_{\nu_i}$ less than $m^{(D)}_{\nu_i}$).

Note that the Majorana mass $M_{\nu_1}$ of the lightest heavy sterile 
neutrino $\nu_{1 R}$ is too light by two orders of magnitude to reach the 
estimated lower bound $M_{\nu_1} \stackrel{>}{\sim} 10^8$ GeV required for 
the working of baryogenesis through thermal leptogenesis [7] (of course, 
in this mechanism $M^{(\nu)*} \neq M^{(\nu)}$ and $M^{(R)*} \neq M^{(R)}$).

In conclusion, our tentative proposal presented here for the Dirac and 
Majorana neutrino mass matrices $M^{(D)}$ and $M^{(R)}$ contains two items: 
({\it i}) the parametrization (20) of Dirac neutrino masses $m^{(D)}_{\nu_i}$ 
in terms of two constants $\mu^{(\nu)}$ and $\varepsilon^{(\nu)}$ determined 
through the conditions (25), and ({\it ii}) the parametrization (32) of 
Majorana neutrino masses $M_{\nu_i}$ by one constant $\zeta$ determined from 
the experimental estimation of solar $\Delta m^2_{21}$. Then, the atmospheric 
$\Delta m^2_{32}$ is {\it predicted} close to its experimental estimate. All 
neutrino seesaw masses $m_{\nu_i}$ and Majorana masses are evaluated. The 
mass spectra $m_{\nu_1} < m_{\nu_2} < m_{\nu_3}$, $m^{(D)}_{\nu_1} < 
m^{(D)}_{\nu_2} < m^{(D)}_{\nu_3}$ and $M_{\nu_1} < M_{\nu_2} < M_{\nu_3}$ 
are {\it hierarchical}, behaving as $1 : 5.4\!\times\! 10^2 : 
3.3\!\times\! 10^3$, $1 : 4.8\!\times\! 10^3 : 8.2\!\times\! 10^4$ and 
$1 : 4.4\!\times\! 10^4 : 2.0\!\times\! 10^6$, respectively, with 
$m_{\nu_1} \sim 1.6\!\times\! 10^{-5}$ eV, $m^{(D)}_{\nu_1} \sim 0.23$ MeV, 
and $M_{\nu_1} \sim 3.3\!\times\! 10^{6}$ GeV.

\vspace{0.2cm}

\vfill\eject

~~~~
\vspace{0.5cm}

{\centerline{\bf References}}

\vspace{0.5cm}

{\everypar={\hangindent=0.6truecm}
\parindent=0pt\frenchspacing

{\everypar={\hangindent=0.6truecm}
\parindent=0pt\frenchspacing

[1]~W. Kr\'{o}likowski, {\it Acta Phys. Pol.} {\bf B 33}, 2559 (2002); 
and references therein. 

\vspace{0.2cm}

[2]~The Particle Data Group, {\it Phys.~Rev} {\bf D 66}, 010001 (2002).

\vspace{0.2cm}

[3]~W. Kr\'{o}likowski, "A proposal of quark mass formula and lepton 
spectrum", to appear in {\it Acta Phys. Pol.} {\bf B}, (2004). 

\vspace{0.2cm}

[4]~{\it Cf. e.g.} V. Barger, D. Marfatia and K. Whisnant, 
{\tt hep--ph/0308123}; and references therein.

\vspace{0.2cm}

[5]~M. Gell-Mann, P. Ramond and R.~Slansky, in {\it Supergravity}, 
edited by F.~van Nieuwenhuizen and D.~Freedman, North Holland, 1979; 
T.~Yanagida, Proc. of the 
{\it Workshop on Unified Theory and the Baryon Number in the Universe}, 
KEK, Japan, 1979; R.N.~Mohapatra and G.~Senjanovi\'{c}, 
{\it Phys. Rev. Lett.} {\bf 44}, 912 (1980).

\vspace{0.2cm}

[6]~W. Kr\'{o}likowski, {\tt hep-ph/0309020}. 

\vspace{0.2cm}

[7]~{\it Cf.} in particular G.-F. Giudice, A. Notari, M. Raidal, A. Riotto 
and A. Strumia {\tt hep--ph/0302092}; T. Hambye, Y. Lin, A. Notari, 
M. Papucci and A. Strumia {\tt hep--ph/0312203}; and references therein.

\vfill\eject

\end{document}